\documentclass[twoside]{siamltex}

\catcode`\@=11
\def\ltsim{\mathrel{\vcenter{\m@th\offinterlineskip
\hbox{$\hfill<\hfill$}\kern.5ex\hbox{$\hfill\sim\hfill$}}}}
\catcode`\@=12
\catcode`\@=12

\usepackage{graphicx}

\begin{document}

\title{Pulse propagation in discrete systems of
coupled excitable cells
\thanks{This research was supported by the DGES grant
PB98-0142-C04-01, by the Third Regional Research Program
of the Autonomous Region of Madrid (Strategic Groups
Action), and by the European Union under grant
RTN2-2001-00349. Received by the editors of SIAM J. Appl.
Math. on  . Manuscript number .}}
\author{A. CARPIO\footnotemark[2]
\and L. L. BONILLA\footnotemark[3] \footnotemark[4]}
\date{June 21, 2001}
\maketitle

\renewcommand{\thefootnote}{\fnsymbol{footnote}}

\footnotetext[2]{Departamento de Matem\'atica Aplicada, Universidad
Complutense de Madrid, 28040 Madrid, Spain. }

\footnotetext[3]{Departamento de Matem\'aticas, Escuela Polit{\'e}cnica
Superior, Universidad Carlos III
de Madrid, Avenida de la Universidad 30, 28911 Legan{\'e}s, Spain.
}
\footnotetext[4]{Also: Unidad Asociada al
Instituto de Ciencia de Materiales de Madrid (CSIC).}

\renewcommand{\thefootnote}{\arabic{footnote}}
\renewcommand{\theequation}{\arabic{section}.\arabic{equation}}
\newcommand{\fin}{\newline \rule{2mm}{2mm}}
\def\RR{\hbox{{\rm I}\kern-.2em\hbox{\rm R}}}
\def\pRR{\hbox{{\tiny \rm I}\kern-.1em\hbox{{\tiny \rm R}}}}
\def\ZZ{\hbox{{\rm Z}\kern-.42em\hbox{\rm Z}}}

\begin{abstract}
Propagation of pulses in myelinated fibers may be
described by appropriate solutions of spatially
discrete FitzHugh-Nagumo systems. In these
systems, propagation failure may occur if either the
coupling between nodes is not strong enough or
the recovery is too fast. We give an asymptotic
construction of pulses for spatially discrete
FitzHugh-Nagumo systems which agrees well
with numerical simulations and discuss evolution
of initial data into pulses and pulse generation at a
boundary. Formulas for the speed and length of pulses are
also obtained.
\end{abstract}



34E15, 92C30. \hspace{2cm} Date: \today

\setcounter{equation}{0}
\section{Introduction}
\label{sec:intro}

Effects of spatial discreteness are important in many
physical and biological systems comprising interacting
smaller components such as atoms, quantum wells, cells,
etc. Examples are the motion of dislocations \cite{fk}, 
and crystal growth and interface motion in crystalline
materials \cite{cah60}, the motion of domain walls in
semiconductor superlattices \cite{bon02,cba01}, sliding of
charge density waves \cite{cdw}, pulse propagation through
myelinated nerves \cite{sleeman}, and so on. The
mathematical study of spatially discrete models is
challenging because of special and poorly understood
phenomena occurring in them that are absent if the
continuum limit of these models is taken. Paramount among
these phenomena is the pinning or propagation failure of
wave fronts in spatially discrete equations. Physically,
the pinning of wave fronts is related to the existence of
Peierls stresses in continuum mechanics \cite{hob65},
relocation of electric field domains \cite{ama01} and
self-sustained oscillations of the current in
semiconductor superlattices \cite{kas97,bon02}, electric
current due to sliding of charge density waves
\cite{cdw}, saltatory propagation of impulses in
myelinated fibers and its failure \cite{sleeman}, etc. 

Mathematical understanding of propagation failure of wave
fronts in spatially discrete equations experienced
significant progress after a paper by Keener \cite{kee87}.
In Ref.~\cite{kee87}, Keener used comparison principles
to characterize pinning of wave fronts and their motion
for spatially discrete reaction diffusion equations of
the form:
\begin{equation}
u_{n,t} = d (u_{n+1}-2u_n+u_{n-1}) + f(u_n),
\label{escalar}
\end{equation}
where $f$ is a bistable source term and $d$ measures
the strength of the coupling. Models described by Eq.\
(\ref{escalar}) include the spatially discrete Nagumo
equation for nerve conduction \cite{fit61,nag62} or
crystal growth \cite{cah60}, and the Frenkel-Kontorova
model for motion of dislocations \cite{fk}. More recently,
a number of results on the existence of traveling wave
fronts $u_n(t)=v(n-ct)$ with smooth profiles $v$ have
been established \cite{zinner,malletparet,bryce}. These
papers do not characterize precisely the propagation
failure of a wave front for critical values of the
control parameters. For a piecewise linear source
function, an explicit description has been given by
F\'ath using extensively the properties of special
functions \cite{fat98}. In the general case, we have
found that propagation failure can be explained as a loss
of continuity of the wave front profile in critical
values of the control parameter \cite{cb01}. Furthermore,
the smoothness of the wave front profile just before the
propagation failure occurs can be exploited to obtain an
analytical description of wave fronts and their speed
near critical parameter values \cite{cb01,siap01}. This
theory has been extended to spatially discrete
reaction-diffusion-convection equations describing
dynamics of domain walls in semiconductor superlattices
in Ref.~\cite{cba01}. 

These advances in the mathematical understanding of
propagation phenomena have occurred for spatially discrete
scalar reaction-diffusion equations. Similar phenomena
occur in models of calcium release at discrete sites
\cite{kee00}. The latter consist of scalar
reaction-diffusion equation with spatially inhomogeneous
source terms that are close to $f(u)$ times a series of
delta functions centered at spatially periodic sites.
Comparatively, little progress has been made
understanding wave propagation and failure in spatially
discrete systems. Anderson and Sleeman \cite{sleeman}
have extended Keener's techniques to discrete
reaction-diffusion systems modelled by the
FitzHugh-Nagumo (FHN) dynamics \cite{fit61,nag62}.
Hastings and Chen \cite{ha99} have proved the existence
of pulse traveling waves for a myelinated nerve model
with a Morris-Lecar type dynamics. They also comment the
difficulties of extending their results to the FHN
system. An attempt to understand the mechanisms of
propagation failure in the FHN system has been carried
out by Booth and Erneux \cite{erneux}. They consider slow
recovery and very special limiting (small) values of the
parameters characterizing the bistable source and the
spatial diffusivity in the FHN system. Furthermore, they
also impose particular boundary and initial conditions.
With these restrictions, they could study how a specific
disturbance localized in one cell propagated to
neighboring ones until the resulting front failed to
propagate. No construction of pulses or formulas for
their velocity were given. 

\begin{figure}
\begin{center}
\includegraphics[width=6cm]{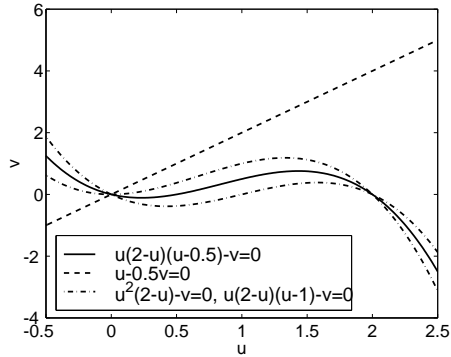}
\caption{Nullclines for the space independent FHN model
with different values of $a$.}
\label{nuliclinas}
\end{center}
\end{figure}

In this paper, we construct asymptotically pulse
solutions of the spatially discrete FHN system describing
nerve conduction through myelinated fibers. We also
discuss how the pulses may fail to propagate. Our ideas
could be extended to spatially discrete systems whose
cell dynamics contains widely separated time scales
corresponding to fast excitation and slow recovery
variables. Among these systems, let us cite models for
bursting behavior in pancreatic $\beta$ cells
\cite{vri98} or the much more difficult case of front
propagation in voltage biased semiconductor superlattices
\cite{bon02}. In the latter, a separation of time scales
exists but it is not obviously included as a small
parameter in the equations. In our presentation, we have
chosen the FHN dynamics for its simplicity. This model
has been widely  used to understand issues that are
obscured by technical complications in more realistic
models of nerve conduction. We consider the following
system of dimensionless equations:
\begin{eqnarray}
\epsilon\, {du_{n}\over dt} &=& d\, (u_{n+1}-2 u_n
+ u_{n-1}) + A u_n (2-u_n)(u_n-a) - v_n , \label{fh1}\\
{dv_{n}\over dt} &=& u_n-Bv_n , \label{fh2}
\end{eqnarray}
$n=0, \pm 1,\ldots$. 
Here $u_n$ and $v_n$ are the membrane potential and the
recovery variable (which acts as an outward ion current)
at the $n$th excitable membrane site (Ranvier node). The
cubic source term is an ionic current and the discrete
diffusive term is proportional to the difference in
internodal currents through a given site. The constants
$A$ and $B$ are selected so that the source terms in the
FHN system are $O(1)$ for $u_n$  and $v_n$ of order 1,
that the only stationary uniform solution is $u_n =
0=v_n$ and the FHN system has excitable dynamics ($A=1$,
$B=0.5$ is a good choice, see Fig.\ \ref{nuliclinas}).
The constant $\epsilon >0$ is the ratio between the
characteristic time scales of both variables. We assume
$\epsilon \ll 1$, that is, fast excitation and slow
recovery. A dimensional version of Eqs.\ (\ref{fh1}) and
(\ref{fh2}) was derived in the Appendix of Ref.~\cite{bc}
from an equivalent-circuit model of myelinated nerves. For
background on similar models, see 
\cite{sco75,str97,kee98,mci99}.  

We shall study pulse propagation in the spatially
discrete FHN system (\ref{fh1}) and (\ref{fh2}) by
asymptotic methods. At first sight, such a task is
hopeless: asymptotic methods require a degree of
smoothness at appropriate time or length scales, and the
spatial variable $n$ in these systems is discrete.
However, we can use the separation between time scales in
the FHN system to show that a pulse is made out of two
`sharp' wave fronts separating regions of slow spatial
variation. Wave fronts are {\em smooth} solutions of the
{\em continuous} variable $z=n-ct/\epsilon$, and
perturbative arguments apply straightforwardly to them.
Thus the theory  of wave front propagation for spatially
discrete scalar reaction-diffusion equations plays an
important role in our construction of pulses. 

\begin{figure}
\begin{center}
\includegraphics[width=10cm]{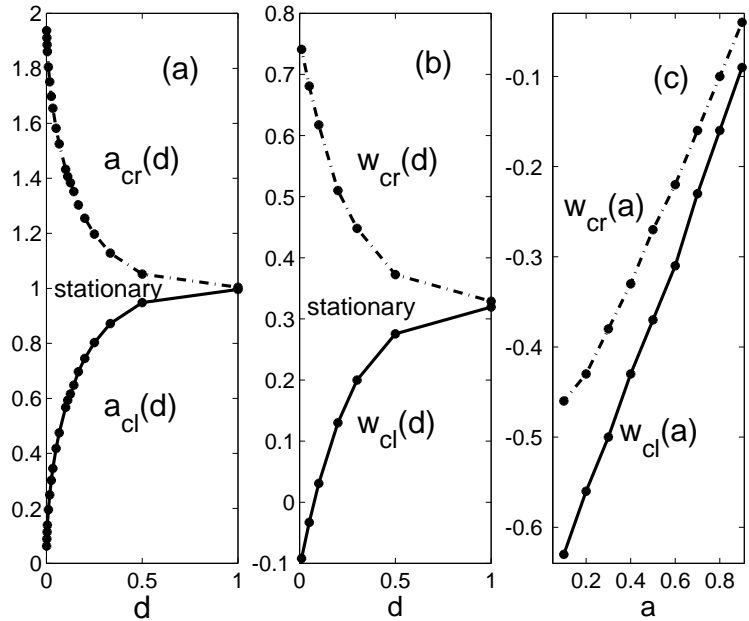}
\caption{(a) Critical values $a_{cl}$ and $a_{cr}$ as
functions of $d$. (b) Critical $w_{cl}(a,d)$ and
$w_{cr}(a,d)$ for $a=0.5$. (c) Critical $w_{cl}(a,d)$ 
and $w_{cr}(a,d)$ for $d=0.5$.}
\label{awcrit}
\end{center}
\end{figure}

Let $U_1< U_2<U_3$ denote the three zeros of the cubic
nonlinearity $f(u)$ in Eq.\ (\ref{escalar}). $U_1$ and
$U_3$ are stable solutions for $d=0$. A wave front is a
solution of Eq.\ (\ref{escalar}) with a smooth profile
$u_n(t)= u(n-ct)$ moving at a speed $c$, such that
$u(\mp\infty) =U_1$ and $u(\pm\infty)=U_3$. If $f(u)$ is
odd about $u=U_2$ and $d$ is sufficiently small, a
stationary solution of Eq.\ (\ref{escalar}) exists and
therefore no wave fronts can propagate \cite{bryce}. As
the source term departs from this symmetric form, front
propagation is made easier. In Ref.~\cite{cb01}, we
selected $d=1$ and $f(u)= F-A g(u)$, where $g(u)$ is odd
about its middle zero and $F$ is an external force that
quantifies departure from symmetry. Notice that we can
obtain Eq.\ (\ref{escalar}) with $d=1/A$ and $f=(F/A) -
g(u)$ after rescaling time. We found that wave fronts
propagate for
$|F|>F_c$, where $F_c>0$ depends on $A$ and the specific
$g(u)$ we adopt. Equivalently, we could set $f(u) = - u
(u-a) (u-2)$ and use $a-1$ as a control parameter instead
of $F$. After the ``external force'' $a$ surpasses a
critical value sufficiently far from the symmetry point
$a=1$, stationary fronts may cease to exist and
propagating wave fronts may appear. See Fig.\
\ref{awcrit}(a). Notice that the limiting case considered
by Booth and Erneux (``slow capture near a limit point''),
$d=O(a^2)$, $a\to 0+$, corresponds to considering the
parameter region in the lowest corner in this Figure. In
this region, propagation failure is the normal situation.
We could alternatively fix $a$ and set $f(u) = - w - u
(u-a) (u-2)$, using $w$ as the control parameter. Then
there are critical values $w_{cl}(a,d)$ and $w_{cr}(a,d)$
such that wave fronts fail to propagate if
$w_{cl}<w<w_{cr}$; see Fig.\ \ref{awcrit}(b). How does
the parameter $a$ affects the critical values of $w$?
Assume that $a<a_{cl}(d)$ for a fixed value of $d$, so
that wave fronts propagate for $w=0$. To compensate this
effect, we need a small critical value $w_{cl}(a,d)$ of
the parameter $w$. As $a$ is departs more and more from
$a_{cl}(d)$, larger and larger critical values
$w_{cl}(a,d)$ are needed to return to the situation of
propagation failure. A similar situation occurs with
$w_{cr}(a,d)$. Thus the critical values of $w$ increase
(in absolute value) as $|a-a_c|$ increases; see 
Fig.~\ref{awcrit}(c). The effect of increasing the
diffusivity $d$ is to shrink the parameter range where
stationary fronts exist. In fact, as $d\to\infty$
(continuum limit), the width of the pinning interval is
conjectured to decrease exponentially fast to zero for
certain nonlinearities \cite{cah60,kin01}. Propagation
failure can be understood as a loss of continuity of the
moving front as appropriate critical parameter values are
approached. Increasing the discrete diffusivity and
deforming the source term sufficiently far from odd
symmetry about its middle zero, both facilitate
propagation of wave fronts \cite{cb01,siap01}. 
Reciprocally, weakening the coupling between cells and
diminishing the ``external force'' $a-1$ helps finding
propagation failure. 

For the spatially discrete FHN system, the description of
wave propagation is more complicated. This also happens
for the spatially continuous FHN system ($D\, \partial^2
u/\partial x^2$ instead of discrete diffusion). Depending
on the initial condition, stable wave trains or pulses
may be approached as time elapses
\cite{rin73,hastings,kee80}. Pulses cannot be obtained
for $\epsilon$ larger than a critical value. For discrete
diffusion, we can construct pulses provided $\epsilon$ is
smaller than a critical value $\epsilon_c(a,d)$, $a$ is
outside a certain interval (corresponding to propagation
failure in the scalar case) and the initial condition is
chosen appropriately. Our construction combines the
theory of front depinning developed in \cite{cb01} with
Keener's asymptotic ideas \cite{kee80} developed for the
FHN model with spatially continuous diffusion. Our results
agree very well with direct numerical solutions of
(\ref{fh1}) and (\ref{fh2}).  

The key ideas of an asymptotic theory for (\ref{fh1}) and
(\ref{fh2}) in the limit as $\epsilon\to 0$ are simple.
First of all, pulses consist of regions where $u_n(t)$
vary smoothly with $n$, separated by moving sharp
interfaces (fronts). In the first type of region, we may
set $\epsilon=d=0$ and obtain a description of slow
recovery. The sharp interfaces are wave fronts with {\em
smooth profiles}, $u_n(t) = u(z)$, $v_n(t) = v(z)$, with
$z=n-ct/\epsilon$. Then $v$ is constant at each side of a
front and the excitation variable $u$ obeys the spatially
discrete Nagumo equation, whose fronts we can
characterize \cite{cb01,siap01}. A stable pulse is
obtained when the velocity of the leading front is equal
to that of the trailing front \cite{kee80}. This
condition fixes the pulse width. Its violation or
propagation failure of any of the fronts bounding the
pulse result in propagation failure thereof. Notice that
we use an analytic expression for the wave front velocity
of the Nagumo equation valid as $a$ is near its critical
value for propagation failure. For small values of $d$,
the front propagation range is narrow and the formula for
wave front velocity holds for all appropriate values of
$a$; see Fig.~\ref{awcrit}(a). For larger values of
$d$, the interval where propagation occurs is wide and we
can use our approximation only for $a$ close to its
critical values $a_{cl}$ and $a_{cr}$. Outside these
parameter ranges, the velocity of the Nagumo wave fronts
should be calculated numerically.

The rest of the paper is organized as follows. In Section
\ref{sec:nagumo}, we recall certain needed results on wave
front propagation and failure for the spatially discrete
scalar reaction-diffusion (Nagumo) equation. Section
\ref{sec:asympt} contains our main theoretical ideas in
this paper, with the asymptotic construction of pulses for
the discrete FHN system. These ideas and our results
are tested by numerically solving the FHN system with
appropriate boundary conditions in Section 
\ref{sec:numerical}. Comments on propagation failure
of pulses in the FHN system are made in Section
\ref{sec:failure}. Section \ref{sec:bdry} briefly
discusses how a pulse may be generated by applying a
temporary stimulus at one end of a fiber with finitely
many nodes. The last Section contains our Conclusions.   

\setcounter{equation}{0}
\section{The spatially discrete Nagumo equation}
\label{sec:nagumo}
We consider the equation
\begin{eqnarray}
{du_{n}\over ds} = d\, (u_{n+1}-2 u_n+u_{n-1})
+ u_n(2-u_n)(u_n- a)-w,  \label{rn}
\end{eqnarray}
for some constant $w$ and denote $h(u,w,a)= u (2-u)
(u-a) - w$. As long as $\hbox{min } h(u,0,a) < w <
\hbox{max } h(u,a,w)$ this is a `cubic' source having
three zeroes $U_i(w,a)$, $i=1,2,3$, $U_1<U_2<U_3$. Wave
front solutions joining $U_1$ and $U_3$ (the two
stable zeros) exist.  A theory of pinning and
propagation of fronts for this type of equations
has been developed in \cite{cb01,cba01}. We
sketch below its implications for (\ref{rn}).

Assume first $w=0$ so that the assymmetry of the
source is controlled by the parameter $a$. For
$d$ fixed, there are values $a_{cl}(d)$ and 
$a_{cr}(d)$ such that:
\begin{itemize}
\item The fronts joining $u=0$ and $u=2$ are stationary
if $a_{cl}(d)\leq a\leq a_{cr}(d)$. No front propagation
is possible.
\item Outside this interval, there exist traveling wave
fronts $u_n(s)= u(n-cs)$ joining $0$ and $2$. For $a>
a_{cr}(d)$, increasing fronts move to the right and
decreasing fronts move to the left. For $a< a_{cl}(d)$,
fronts move in the opposite way: decreasing fronts move
to the right and increasing fronts move to the left.
\end{itemize}
The values $a_{cl}(d)$ and $a_{cr}(d)$ can be
approximately calculated as follows. In a large
lattice, we decrease or increase $a$ from $1$ till we
obtain an stationary solution $u_n(a)$ whose linear
stability problem has a zero eigenvalue. See Fig.\
\ref{awcrit}.

Now we fix $a$ and vary $w$. The asymmetry of the
source is  controlled by $a$ and $w$. For fixed $d$ and
$a$, critical values $w_{cl}(a,d)$ and $w_{cr}(a,d)$ are
found such that:
\begin{itemize}
\item The fronts joining $U_1(w,a)$ and
$U_3(w,a)$ are stationary if $w_{cl}(a,d) \leq w \leq
w_{cr}(a,d)$.
\item Outside this interval, there exist traveling
wavefronts $u_n(s)=u(n-cs)$ joining $U_1(w,a)$ and
$U_3(w,a)$. For $w<w_{cl}$, these fronts move to the
left if they increase from $U_1$ to $U_3$, and to the
right if they decrease from $U_3$ to $U_1$. For $w>
w_{cr}$, fronts decreasing from $U_3$ to $U_1$ move to
the left, and increasing fronts move to the right.
\end{itemize}
To calculate $w_{cl}$ and $w_{cr}$, we start by fixing
$a$ and finding a value, $w=w_0$, at which stationary
solutions exist for a large lattice. We now decrease
or increase $w$ from this value till we obtain a
stationary solution $u_n(w)$ whose linear stability
problem has a zero eigenvalue. See Fig.\
\ref{awcrit}.

For $w$ near any of its critical values, we can use the
following formula to predict the speed of the fronts for
$|w|> |w_c|$: 
\begin{eqnarray}
c(a,d,w) \sim \mbox{sign}(w-w_c) {\sqrt{\alpha \beta
(w-w_c)} \over \pi}\,. \label{speed}
\end{eqnarray}
The parameters $\alpha$ and $\beta$, given by $\alpha=\sum
\phi_n$, $\beta={1\over 2} \sum [-6 u_n(w_c)+2(2+a)]
\phi_n^3$ \cite{cb01,siap01}, are functions of
$a$, $d$ and the critical value of $w$. In these
formulas, $\phi$ is a positive eigenfunction of the
linear stability problem for $u_n(w_c)$ with $\sum
\phi_n^2 =1$ and $u_n(w_c)$ is a stationary solution of
(\ref{rn}) with $w=w_c$ \cite{siap01}. If $w$ is not close
to its critical values, the speed $c(a,d,w)$ should be
calculated numerically.

A peculiarity of the Nagumo equation is the scenario for
front propagation failure. As we approach the critical
values for $a$, $w$, or any other appropriate pinning
control parameter, the front profiles become less smooth
and a number of steps  appear. In the limit as the control
parameter tends to its critical value, the transition
regions between steps become infinitely steep, the front
profile becomes discontinuous and its velocity vanishes
\cite{cb01,siap01}.

\setcounter{equation}{0}
\section{Asymptotic construction of pulses}
\label{sec:asympt}
As we will discuss below, an appropriate initial condition
evolves towards a pulse. In particular, we need
to fix the parameters $d>0$, $a < a_{cl}(d)$ (the case $a
> a_{cr}(d)$ follows by symmetry) and $\epsilon$ smaller
than a certain critical value, $\epsilon_c(a,d)$. This
latter condition also holds for the spatially continuous
FHN system which has two pulse solutions (one stable and
one unstable) for $\epsilon<\epsilon_c$. These solutions
coalesce at $\epsilon_c$ and cease to exist for larger
$\epsilon$ \cite{nag62,rin73}. A pulse consists of regions
of smooth variation of $u$ on the time scale $t$,
separated by sharp interfaces in which $u$ varies rapidly
on the time scale $T=t/\epsilon$. In the regions where
$u$ varies smoothly, we can set $\epsilon= d=0$, thereby
obtaining the reduced problem,
\begin{eqnarray}
u_n (2-u_n)(a-u_n) - v_n = 0, \label{slow1}\\
{dv_{n}\over dt} = u_n - B\, v_n. \label{slow2}
\end{eqnarray}
These regions are separated by sharp interfaces (moving
fronts) at which $u_n$ varies rapidly as $u_n(t) = u(z)$, $v_n(t) = v(z)$, with
$z=n-ct/\epsilon$. There, to leading order,
\begin{eqnarray}
- c\, {du\over dz} &=& d\, [u(z+1)-2u(z)+u(z-1)]
+ u(z)\, (2-u(z))\, (a-u(z)) - v, \label{fast1} \\
- c\, {dv\over dz} &=& 0. \label{fast2}
\end{eqnarray}
Thus $v$ is a constant equal to the value $v_n(t)$ at
the last point in the region of smooth variation before
the front. Equation (\ref{fast1}) has a wave front
solution as discussed in the previous Section. We can
now discuss different regions in the asymptotic
description of a pulse:

\begin{enumerate}
\item The region of smooth variation of $u$ in
front of the pulse, described by (\ref{slow1}) and
(\ref{slow2}). In this region, $u_n = U_1(v_n)$, so that
$${dv_{n}\over dt} = U_1(v_n) - B\, v_n ,
$$
and initial data evolve exponentially fast towards
equilibrium, $u_n = v_n = 0$.

\item The pulse leading edge. Let $v(t)$ be the value of
$v_n$ at the last point of the region in front of the
pulse. Eventually, $v\to 0$. At the leading edge,
$u_n(t) = u(n-ct/\epsilon)$ is a wave front moving
towards the right with speed $C= c(a,d,v)/\epsilon$
measured in points per unit time $t$. We have the
boundary conditions $u(-\infty) = U_3(v)$ and $u(\infty)
= U_1(v)$ for the monotone decreasing profile $u(z)$
which satisfies (\ref{fast1}). It is convenient to call
$c_-(v)= c(a,d,v)$. Eventually, $C\sim c_-(0)/\epsilon$,
and $u_n$ decreases from $u_n=2$ to $u_n =0$ across the
leading edge of the pulse.

\item Region between fronts: $u_n=U_3(v_n)$
and
$${dv_{n}\over dt} = U_3(v_n) - B\, v_n .
$$
There is a finite number of points in this region. On
its far right, $v_n=v\to 0$. As we move towards the
left, $v_n$ increases until it reaches a certain value
$V(t)$ corresponding to that in the trailing wave front.

\item Trailing wave front: $v_n(t)=v(z)=V$, and $u_n(t)
=u(z)$ obeys (\ref{fast1}) with boundary conditions
$u(-\infty)= U_1(V)$ and $u(\infty)= U_3(V)$. This front
increases monotonically with $z$ and it moves with speed
$C= c(a,d,V)/\epsilon$ measured in points per unit time
$t$. It is convenient to denote $c_+(V)=c(a,d,V)$. We
shall indicate how to determine $V$ below. Clearly, if
the pulse is to move rigidly, we should have $c_+(V)=
c_-(0)$ after a sufficiently long transient period.

\item Pulse tail. Again $u_n=U_1(v_n)$ and $dv_{n}/dt=
U_1(v_n)-Bv_n$. Sufficiently far to the left, $v_n=u_n=
0$.
\end{enumerate}

The number of points between wave fronts of the pulse is
not arbitrary: it can be calculated following an
argument due to Keener for the spatially continuous case
\cite{kee80}. Let $\tau$ be the delay between fronts,
i.e. the time elapsed from the instant at which the
leading front traverses the point $n=N$ to the instant
when the trailing front is at $n=N$. Clearly,
\begin{eqnarray}
\tau =\int_{v(t-\tau)}^{V(t)} {dv \over U_3(v)-Bv}\,.
\label{tau}
\end{eqnarray}
The number of points between fronts, $l(t)$, can be
calculated as
\begin{eqnarray}
l ={1\over\epsilon}\,\int_{t-\tau}^{t} c_-(v(t))\, dt\,.
\label{l}
\end{eqnarray}
On the other hand, the separation between fronts
satisfies the equation
\begin{eqnarray}
{dl\over dt} = {c_{-}(v(t)) - c_{+}(V(t))\over \epsilon}
\,.   \label{eql}
\end{eqnarray}
The three equations (\ref{tau}), (\ref{l}) and
(\ref{eql}) can be solved to obtain the three unknowns
$\tau$, $l$ and $V(t)$. (The function $v(t)$ is
determined by solving (\ref{slow2}) with $u_n= U_1(v_n)$
in the region to the left of the leading front).

After a transient period, $v(t)\to 0$ and $V(t)\to
V$ (a constant value), so that we have the simpler
expressions
\begin{eqnarray}
\tau =\int_{0}^{V} {dv \over U_3(v)-Bv}\,,
\label{tau1}\\
{dl\over dt} = {c_{-}(0) - c_{+}(V)\over \epsilon}\,,
\label{eql1}
\end{eqnarray}
instead of (\ref{tau}) and (\ref{eql}), respectively.
The number of points at the pulse top is now
\begin{eqnarray}
l= {c_-(0)\tau\over\epsilon} = {c_-(0)\over\epsilon}\,
\int_{0}^{V} {dv \over U_3(v)-Bv}\,. \label{l1}
\end{eqnarray}
This equation yields $V$ as a function of $l$. Then
(\ref{eql1}) becomes an autonomous differential equation
for $l$ that has a stable constant solution at $l=l^*$
such that $c_-(0)= c_+(V(l))$: At $l=l^*$, the right
hand side of (\ref{eql1}) has a slope $-[U_3(V)-BV]\,
c'_+(V)/ c_-(0) <0$.

Recapitulating, for appropriate initial conditions,
leading and trailing fronts of a pulse evolve until $l$
reaches its stable value at which $c_-(0)= c_+(V(l^*))$
and (\ref{l1}) holds. To compute $l^*$, we first
determine $V^*=V(l^*)$ by using $c_-(0) =c_+(V(l^*))$.
Then we calculate $\tau=\tau^*$ (which does not depend on
$\epsilon$!) from (\ref{tau1}) and $l^*=c_-(0)\tau^*/
\epsilon$. Our construction breaks down if the number of
points between fronts falls below 1. This yields an
upper bound for the critical value of $\epsilon$ above
which pulse propagation fails: $\epsilon_c \sim c_-(0)
\tau^*$.

The asymptotic length of the pulse tail is obtained by
first calculating the time needed for $v_n$ to go from
$0$ to $V(l^*)$ to the left of the trailing front:
$T =\int_0^{V} dv /[U_1(v)-v]$. The tail length is then
$L= c_-(0)T/\epsilon$.

\setcounter{equation}{0}
\section{Numerically calculated pulses}
\label{sec:numerical}
\begin{figure}
\begin{center}
\includegraphics[width=10cm]{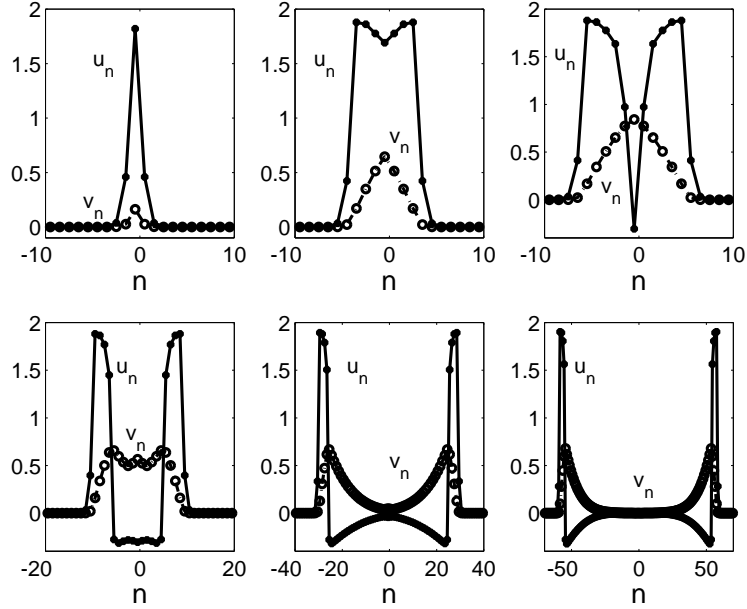}
\caption{Splitting of an initial profile into two pulses
propagating in opposite directions for $d=0.1$, $a=0.5$
and $\epsilon=0.006$.}
\label{split}
\end{center}
\end{figure}

We shall compare numerical solutions for different
representative values of $d$ with the approximate pulses
provided by our theory. As initial data, we have adopted
our approximate pulses. We have also used hump--like
profiles with compact support for $u_n(0)$ and $v_n(0)$.
It is important that $v_n(0) =0$ at the leading edge of
$u_n(0)$ and to its right, and that $v_n(0)= V\approx
w_{cr}(d)$ at the trailing edge, where $v_n(0)$ reaches
its maximum. Had we chosen $v_n(0)=0$ for all $n$, the
$u_n$ profile would split in two pulses traveling in
opposite directions as time elapses; see
Fig. \ref{split}. The region between leading and
trailing fronts of the pulse acquires its asymptotic
shape quite fast, but the pulse tail is usually rather
long and evolves slowly towards its final form.

\begin{figure}
\begin{center}
\includegraphics[width=10cm]{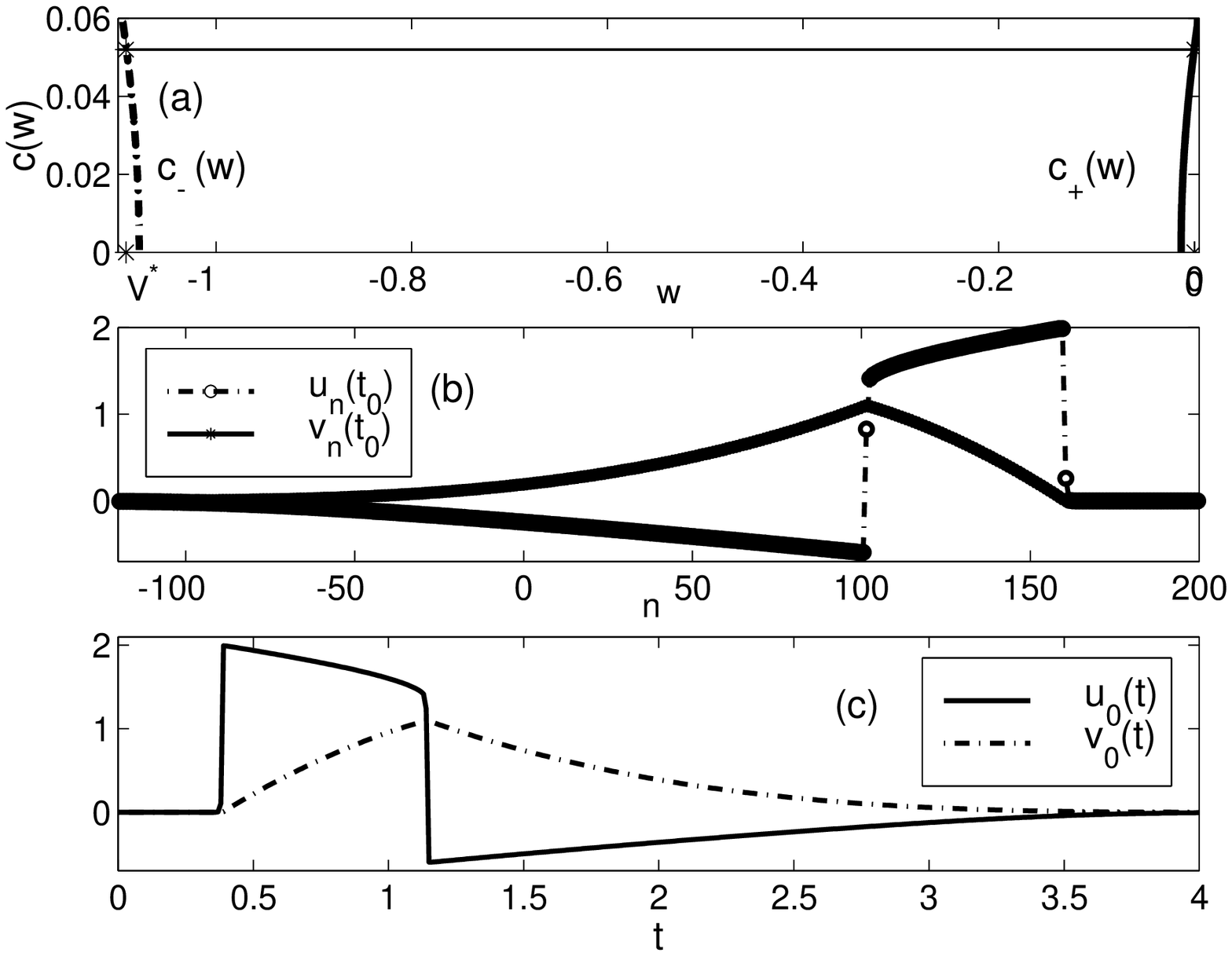}
\caption{(a) Predicted speeds for the Nagumo equation
(\ref{rn}) with $d=0.01$ and $a=0.1$. (b) FHN pulse for
$\epsilon=0.001$ (c) Trajectory of one point, $u_0(t)$, as
the FHN pulse propagates through it.}
\label{d001}
\end{center}
\end{figure}

\begin{itemize}

\item $d=1$. In this case, $a_{cl}=0.996$. We choose
$a=0.99 < a_{cl}$ and analyze first front propagation for
the rescaled Nagumo equation (\ref{rn}). The front
propagation thresholds for these values of $d$ and $a$
are $w_{cl}=0.0038$ and $w_{cr}=0.0095$. Fig.\
\ref{d1}(a) shows the speeds of leading and trailing
fronts as functions of $w$, as predicted by Eq.\
(\ref{speed}). For $w=0$, the leading front should move
at speed $c_-(0)=0.0093$. The relation $c_+(V)=c_-(0)$
yields the asymptotic value $V^*=0.0133$ at the trailing
front joining $U_1(V^*)= -0.00665$ to $U_3(V^*)=1.9933$.
The time elapsed between fronts is $\tau^*=0.00652$, as
calculated from (\ref{tau1}). Then our upper bound for
the critical value of $\epsilon$ is $\epsilon_c =
0.000064$. Choosing a smaller value, $\epsilon=
0.000005$, we obtain a pulse speed $C=c_-(0)/\epsilon
=1869$ points per unit time and a pulse width of $l^*= C
\tau^*\sim 13$ points. Our numerical solution of the full
FitzHugh-Nagumo system (\ref{fh1}) and (\ref{fh2}) yields
a pulse speed $C= 2000$ and a width of 13 points for
$\epsilon=0.000005$. The trailing front joins $-0.006647$
and $1.993$ with
$V^*=-0.0133$. See Fig.\ \ref{d1}(b). Note that the
relative error in the predicted speed $C$ is $0.0655$.
Obviously, rescaling the speed to $C=c_-(0)/\epsilon$
amplifies the error in our predictions. We have not been
able to observe pulses for $\epsilon \geq 0.0000076$ which
is smaller but not far from our estimation $\epsilon_c =
0.000064$.

\begin{figure}
\begin{center}
\includegraphics[width=10cm]{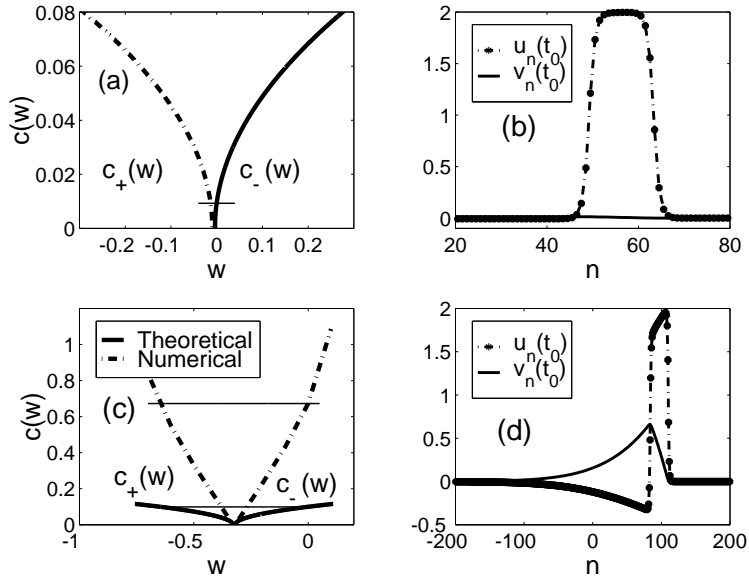}
\caption{(a) Predicted speeds for wave fronts of the
Nagumo equation (\ref{rn}) with $d=1$, $a=0.99$. The
horizontal line marks the condition $c_+(w)=c_-(0)$,
thereby graphically yielding $w=V^*$. (b) FHN pulse for
$\epsilon=0.000005$. (c) Predicted and numerical speeds
for wave fronts of Eq.\ (\ref{rn}) with $d=1$, $a=0.5$.
The horizontal lines mark $c_+(w)=c_-(0)$. (d) FHN pulse
for
$\epsilon=0.01$.}
\label{d1}
\end{center}
\end{figure}

Let us choose now $a=0.5$ which is far from $a_{cl}$.
Then, $w_{cl}=0.3194$ and $w_{cr}=0.3287$. Eq.\
(\ref{speed}) predicts $c_-(0)= 0.09983$ whereas the
trailing front joins $-0.316$ to $1.71$ at $V^*=0.6$. If
$\epsilon=0.01$, the speed and width of the pulse are
$C=9.983$ and $l^*=0.351 C\sim 4$, according to our
theory. Numerically, we observe $C=64.7$ and $l^*=25$. The
source of these large errors is the value $c_-(0)=
0.09983$ predicted with formula (\ref{speed}). If we
replace this value by the numerical front speed
calculated directly, $c_-(0)=0.673$, we obtain $C=67.3$
and $l^*\sim 24$ points, which fit  better the numerically
observed values.

\item $d=0.1$. In this case, $a_{cl}=0.567$ and we shall
choose $a=0.5 < a_{cl}$. Let us first analyze front
propagation for the rescaled Nagumo equation (\ref{rn}).
For these values of $d$ and $a$, we obtain $w_{cl}=
0.0307$ and $w_{cr}=0.6175$. Fig.\ \ref{d01}(a) shows
the predicted speeds of leading and trailing fronts as
functions of $v$, as given by formula (\ref{speed}). For
$w=0$, the leading front should move with speed $c_-(0)=
0.075662$. At the trailing front, $c_+(V)=c_-(0)$ yields
$V^*=0.648$ and $U_1(V^*)=-0.33328$ and $U_3(V^*) =
1.666$. The time elapsed between fronts is $\tau^*=
0.39266$, which gives $l^*= 0.0297/\epsilon$. Our bound
for the critical value of $\epsilon$ is $\epsilon_c=
c_-(0)\tau^* = 0.029$. Selecting $\epsilon=0.003$, we
predict $C= 25.22$ and $l^*\sim 10$ points. Direct
numerical calculations yield a pulse speed $C=26.38$ and
a pulse width about $10$ points. The trailing front joins
$-0.3269$ to $1.675$ with $V^*=0.6578$. See Fig.\
\ref{d01}(b). We have not been able to obtain pulses
for $\epsilon\geq 0.007$, which is four times smaller than
our upper bound of 0.029.

\begin{figure}
\begin{center}
\includegraphics[width=10cm]{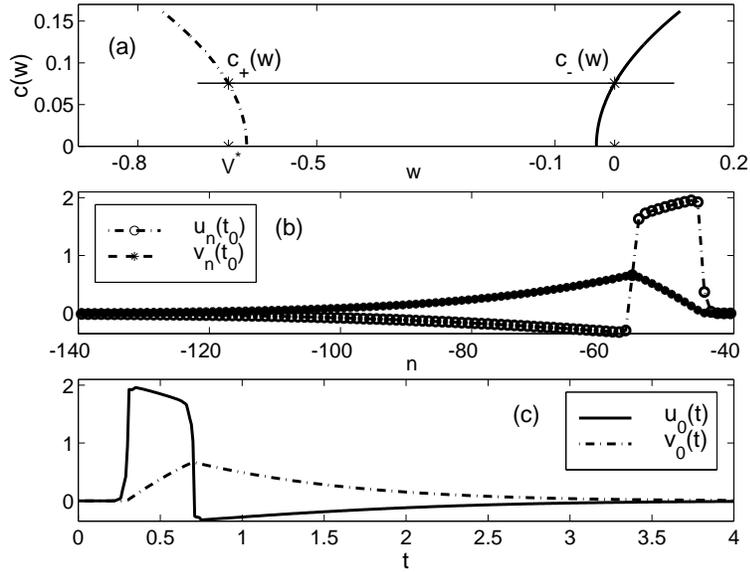}
\caption{(a) Predicted speeds for the Nagumo equation
(\ref{rn}) with $d=0.1$ and $a=0.5$. The horizontal line
graphically yields $V^*$ such that $c_+(V^*)=c_-(0)$. (b)
Profiles of the FHN pulse for $\epsilon=0.003$ (c)
Trajectories of one point, $u_0(t)$, $v_0(t)$, as the FHN
pulse propagates through it.}
\label{d01}
\end{center}
\end{figure}

\item $d=0.01$. In this case, $a_{cl}=0.195$ and we shall
choose $a=0.1 < a_{cl}$. Let us analyze first front
propagation for the rescaled Nagumo equation (\ref{rn}).
The front propagation thresholds for these values of $d$
and $a$ are $w_{cl}=0.0136$ and $w_{cr}=1.0784$. Fig.\
\ref{d001}(a) shows the predicted speeds of leading and
trailing fronts as functions of $w$ according to Eq.\
(\ref{speed}). For $w=0$, the leading front should move
with speed $c_-(0)=0.052$. Then the trailing front has
have $V^*=1.092$ corresponding to $c_+(V)=c_-(0)$, and
it joins $U_1(V^*)=-0.6$ to $U_3(V^*)=1.4$. The time
elapsed between fronts is $\tau^*=0.748$ and the pulse
width, $l^*=0.297/\epsilon$. Our bound for the critical
value of $\epsilon$ is $\epsilon_c= c_-(0)\tau^* =
0.058$. Selecting $\epsilon=0.001$, we predict $C= 52$
and $l^*=39$ points. Numerical observations yield $C=
77.7$ (a relative error of $0.3$) and a pulse width of
$59$ points. Furthermore, the trailing front joins
$-0.59$ to $1.4$ with $V^*=1.095$. See Fig.\
\ref{d001}(b). Again the observed errors in the pulse
speed and width are due to errors in the prediction of
$c_-(0)$ given by formula (\ref{speed}). Replacing this
value by the numerically computed front speed $c_+(0)=
0.078$, we obtain $C=78$ and $l^*\sim 58$, better fit to
the real values.
\end{itemize}

Let us now describe the situation for other values of
$d$. Our asymptotic theory agrees with the numerical
results provided $\epsilon$ is sufficiently small, but
the velocity of the Nagumo wave fronts should be either
approximated by Eq.\ (\ref{speed}) or calculated
numerically depending on how close to zero $w_{cl}$ 
happens to be.  For $d<0.01$, the length of the intervals
where fronts of the Nagumo equation propagate is very
small. Then the front speeds are always very small and
given by Eq.\ (\ref{speed}) with great accuracy. Our
asymptotic description of the pulse agrees very well with
numerical solutions of the FHN system. If $d>1$, the
spatially discrete FHN system can be approximated by its
continuum limit. The length of the pinning intervals for
the Nagumo equation is below 0.001, and the wave front
velocities are essentially a correction of the wave front
velocities for the spatially continuous Nagumo equation
\cite{kee98}:
\begin{eqnarray}
c = \sqrt{d}\, c_0 \,\left(1 - {k_1 c_0^2\over 2d}+
O\left({c_0^4\over d^2}\right) \right)\,,\label{c1}\\
c_0 = {2U_2(w) - U_1(w)-U_3(w)\over \sqrt{2}}\,,
\label{c2}\\
k_1 = - {\int_{-\infty}^\infty e^{-c_0^2 s} V'_0(s)
V''''_0(s)\, ds \over 12 c_0^4\int_{-\infty}^\infty
e^{-c_0^2 s} V'_0(s) V''_0(s)\, ds}\,. \label{c3}
\end{eqnarray}
Here $V_0$ is the appropriate wave front solution of the
equation
\begin{eqnarray}
c_0^{-2}\, V''_0 - V'_0 + V_0\, (2-V_0)(V_0-a) - w =0 .
\label{c4}
\end{eqnarray}

\setcounter{equation}{0}
\section{Propagation failure}
\label{sec:failure}
Two facts may lead to propagation failure: a value of
$\epsilon$ that is too large or $a\in (a_{cl}(d),
a_{cr}(d))$. 

\begin{figure}
\begin{center}
\includegraphics[width=10cm]{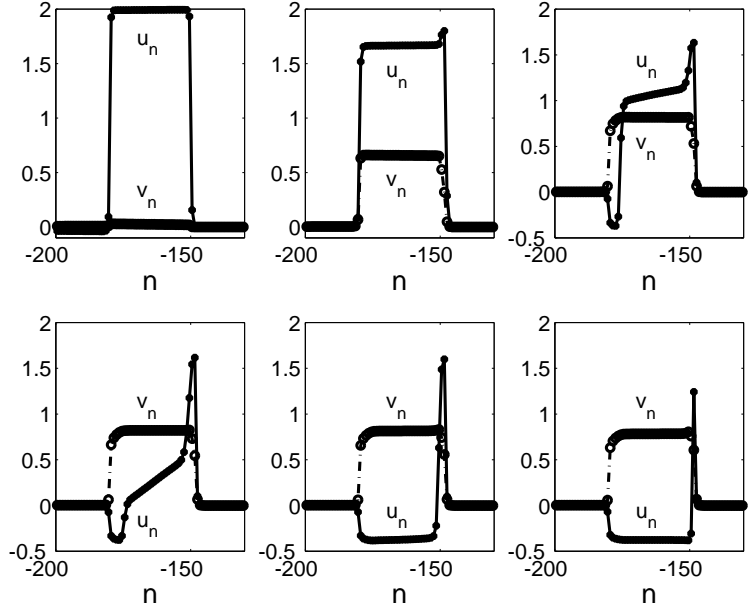}
\caption{Snapshots of the excitation and recovery
variables for $d=0.1$, $a=0.5$ and $\epsilon=0.007$
illustrating propagation failure of the pulse for
$\epsilon>\epsilon_c$.}
\label{failure}
\end{center}
\end{figure}

\begin{figure}
\begin{center}
\includegraphics[width=10cm]{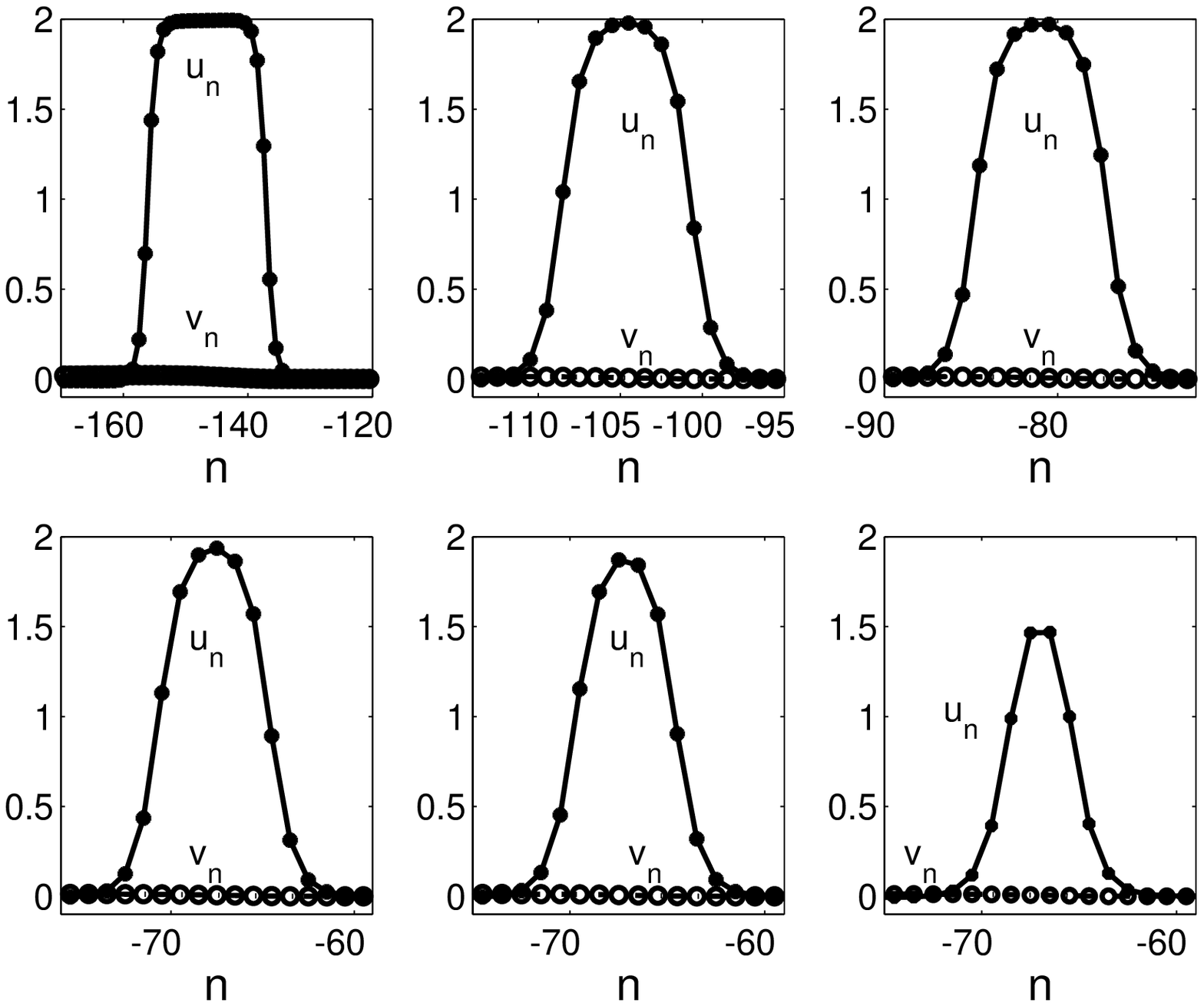}
\caption{Same as Figure \ref{failure} for $d=1$, $a=0.99$
and $\epsilon=0.0000076$.}
\label{failure2}
\end{center}
\end{figure}

Let us consider the first cause of propagation failure
now. If $\epsilon$ surpasses a certain critical value
$\epsilon_c$, recovery is too fast and a stable pulse
cannot be sustained. This situation also occurs in
spatially continuous FHN systems. In these systems, there
exist two pulses (one pulse is stable, the other unstable)
for $\epsilon <\epsilon_c$, they coalesce at $\epsilon_c$
and cease to exist for larger $\epsilon$. In the discrete
FHN system, the phenomenon of wave front propagation
failure implies that pulses may propagate only if $a<
a_{cl}(d)$ or $a> a_{cr}(d)$. As indicated by Eq.\
(\ref{l1}), the number of points between the two fronts of
the stable pulse decreases as $\epsilon$ increases
towards $\epsilon_c(a,d)$. Eventually the two fronts
coalesce and it is not possible to propagate a stable
pulse for $\epsilon >\epsilon_c(a,d)$. If we start with an
appropriate pulse-like initial condition, we find the
scenario of propagation failure depicted in
Figs.~\ref{failure} and \ref{failure2}. For small $d$
($d=0.1$), the variable $v_n$ ceases to be almost
constant at the leading edge of the pulse and the
distance between the two fronts diminishes. While
$v_n\sim 0$ at the rightmost point of the leading front,
$v_n\sim w>0$ at the leftmost point. Thus, $u_n$ in this
front decreases from $U_3(w)$ to zero as $n$ increases.
The value $w$ increases with $\epsilon$ and $U_3(w)$
decreases. At the same time, the leading front speed
diminishes as $w$ increases till $w$ surpasses the
propagation threshold and the leading front stops. Since
the back front goes on moving, the pulse vanishes, see
Fig.\ \ref{failure}. For large $d$ ($d=1$), a decremental
pulse is formed. Its width and height decrease as it
moves until it disappears, see Fig.\
\ref{failure2}. Numerical simulations of the FHN system
show that $\epsilon_c\to 0$ as $a$ tends to either
$a_{cl}(d)$ or $a_{cr}(d)$.

Let us assume now that $a\in (a_{cl}(d),a_{cr}(d))$.
Then, the leading front cannot propagate with $v_n = v =
0$. We need $v_n \sim v < w_{cl}(a,d) <0$. However, in
the region in front of the leading edge, $v_n$ and $u_n$
evolve towards 0, whereas we have $u_n>0$ at the leading
front. Thus $d v_n/dt = u_n -B v_n \geq 0$ there and
$v_n$ will increase until $v_n>0$, which contradicts our
previous assumption. Thus we cannot have stable
propagating pulses. Furthermore, there are no stationary
pulses of the type we have discussed for this range of
$a$: if
$v_n=u_n/B$, the source $u_n(2-u_n)(a-u_n)-v_n =
u_n(2-u_n)(u_n-a)- u_n/B$ has only one zero, not three as
in our construction. This does not preclude existence of
other pulses, such as those corresponding to the homoclinic
orbit in the phase space of the spatially continuous FHN
system. However, we have not observed stable stationary
pulses of this type in the spatially discrete FHN system.

\setcounter{equation}{0}
\section{Pulse generation at a boundary}
\label{sec:bdry}
So far, we have considered the motion of a pulse (or its
failure) in a sufficiently large myelinated nerve fiber.
We have not discussed how such a pulse might be created
in a more realistic situation. Clearly, nerve fibers have
finitely many Ranvier nodes and pulses are typically
generated at the fiber boundary. Thus we are led to
consider how a pulse might be generated by an excitation
at a boundary and how the pulse propagates or fails to in
a finite fiber. This problem was tackled by Booth and
Erneux \cite{erneux} using parameter values for which the
FHN pulse fails to propagate. We shall now discuss
different parameter ranges.

\begin{figure}
\begin{center}
\includegraphics[width=10cm]{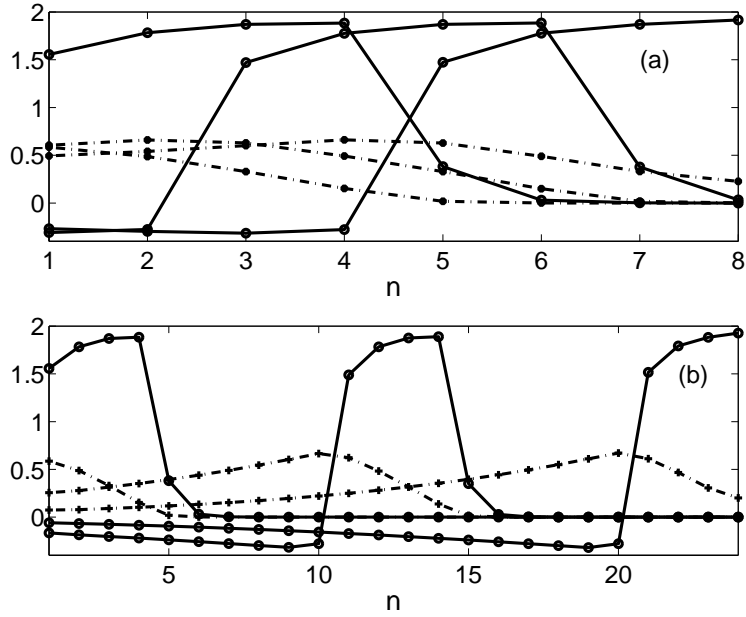}
\caption{Snapshots of the excitation (solid line) and
recovery (dotted line) variables for a FNH system with
$N$ nodes and $d=0.1$, $a=0.5$ and $\epsilon=0.006$. (a)
Profiles at times 0.4, 0.6 and 0.8 for $N=8$. (b)
Profiles at times 0.4, 1.4 and 2.4 for $N=24$. }
\label{bfh}
\end{center}
\end{figure}

\begin{figure}
\begin{center}
\includegraphics[width=10cm]{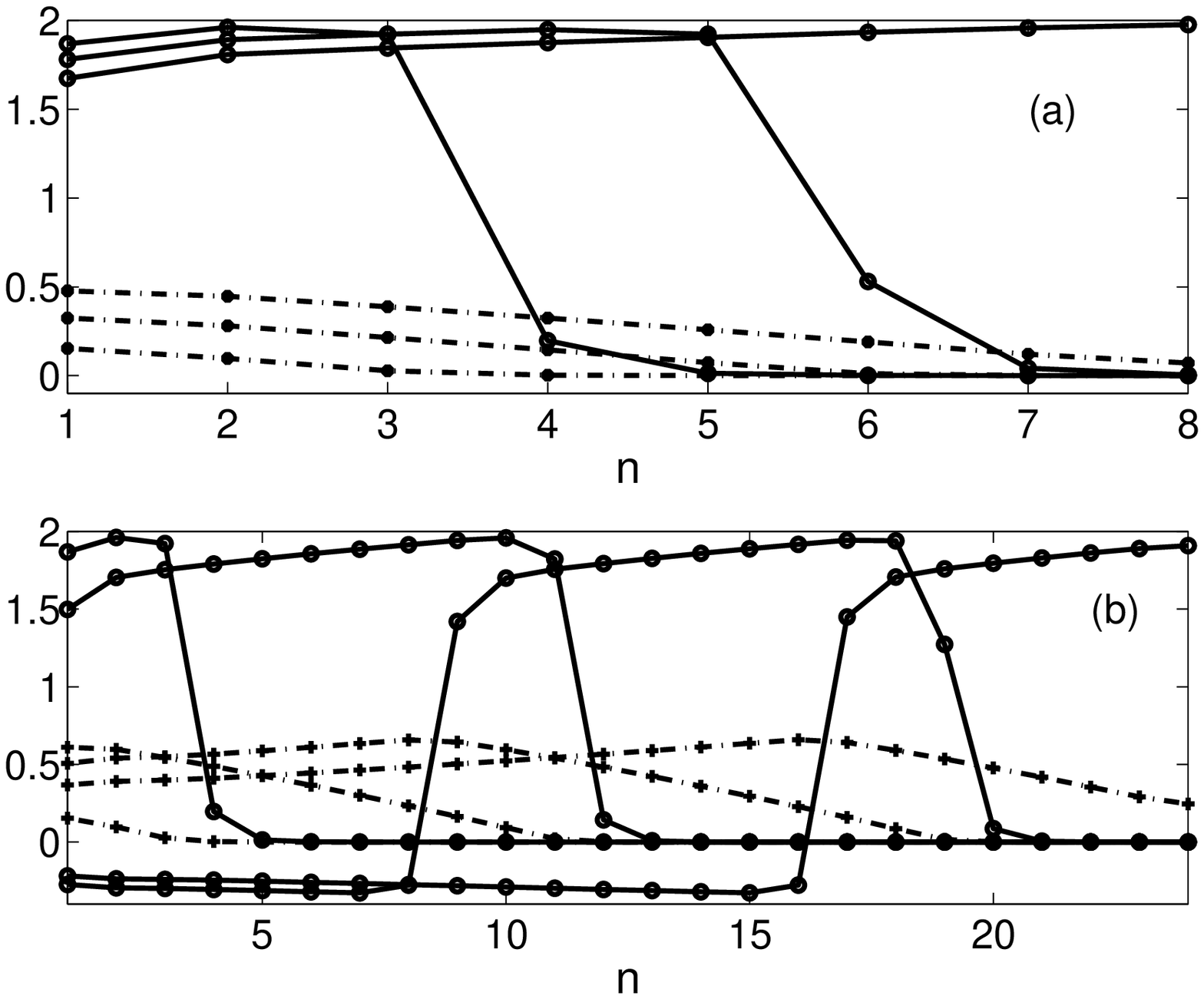}
\caption{Snapshots of the excitation (solid line) and
recovery (dotted line) variables for a FNH system with
$N$ nodes and $d=0.1$, $a=0.5$ and $\epsilon=0.003$. (a)
Profiles at times 0.1, 0.2 and 0.3 for $N=8$. (b)
Profiles at times 0.1, 0.4, 0.7 and 1.0 for $N=24$.}
\label{bfh2}
\end{center}
\end{figure}

Nerve fibers may have either a few Ranvier nodes (e.g., 20
for the central nervous system \cite{mci99}) or several
hundred nodes (for the peripheral nervous system
\cite{str97}). Thus we shall consider a finite FHN system
with $N$ nodes and a Neumann boundary condition at the
right end, $u_{N+1}=u_{N}$. At the left end, we impose
$u_{0}(t)=2$ for $0\leq t\leq 0.05$, and $u_0(t)=0$ for
$t>0.05$. The results corresponding to parameter values
$d=0.1$ and $a=0.5$ are depicted in Figures \ref{bfh} (for
which $\epsilon=0.006$) and \ref{bfh2} (for which
$\epsilon =0.003$). The asymptotic theory predicts that 
fully developed FHN pulses (corresponding to $N=\infty$)
would have widths of $l^*\approx 5$ and $l^*\approx 10$,
respectively. The left boundary condition ensures that
the membrane potential $u_n$ is excited during sufficient
time, so that a wave is generated at the left end of the
fiber. 

The excitation at the left boundary induces a wave front
that propagates with velocity approximately given by
$C= c_-(0)/\epsilon$ along the finite fiber for the
parameter values we consider. For example, $C\approx
12.6$ for $\epsilon=0.006$, which is close to the
numerically observed value of 10 in Fig.~\ref{bfh}.
Similarly, $C\approx 25.22$ for $\epsilon=0.003$, which is
close to the numerically observed value of 26 in
Fig.~\ref{bfh2}. If the fiber is long enough, a second
wave front follows the first one and their mutual
distance rapidly approaches the asymptotic value $l^*$
(the number of nodes between fronts is 4 in Fig.~\ref{bfh}
while the asymptotic theory predicts $l^*\approx 5$; in
Fig.~\ref{bfh2}, numerical observation confirms the
asymptotic value $l^*\approx 10$). The numerical solution
of the finite FHN system shows that an eventually
truncated FHN pulse comprising the two wave fronts and
the region between them is formed provided $N$ is at
least twice $l^*$. Otherwise at best only the first wave
front is shed at the boundary, as shown in
Fig.~\ref{bfh2}(a). Pulses fail to propagate in fibers
whose parameters fall in the propagation failure region,
as discussed in Section \ref{sec:failure}. 

\setcounter{equation}{0}
\section{ Conclusions}
\label{sec:conclusions}
We have constructed stable pulses of the spatially
discrete FitzHugh-Nagumo system by asymptotic methods. In
a pulse, there are regions where the excitation variable
varies smoothly separated by sharp fronts. These fronts
are solutions of the discrete Nagumo equation with a
constant value of the recovery variable. Their shape and
speed can be calculated approximately near parameter
values corresponding to front propagation failure or near
the continuum limit. For long times, their width is given
by the only stable solution of a one-dimensional
autonomous system. We have compared the asymptotic
results with numerical solutions of the FHN system and
analyzed different scenarios for failure of pulse
propagation. Besides the classical scenario of scarce
separation between time scales of excitation and recovery
(large $\epsilon$ as in the spatially continuous FHN
system), propagation failure of fronts for the spatially
discrete Nagumo equation provides a different mechanism of
pulse propagation failure. Wave fronts and pulses can be
generated at a boundary and propagate or fail to
propagate along a finite FHN system. If the number of
nodes is sufficiently large, the two wave fronts
comprising a FHN pulse can be shed at the boundary and
their separation rapidly reaches the value given by the
asymptotic theory. This is so even if the fiber is too
short to accommodate the slowly varying regions at the
back of the second wave front of the pulse. In long
enough fibers, a fully developed FNH pulse may be
generated by an overthreshold stimulus applied during a
short time at one end of the fiber.



\end{document}